# Perturbational perspective on Kekulè valence structures of biphenylene and related hydrocarbons

Viktorija Gineityte

Institute of Theoretical Physics and Astronomy, Vilnius University, Sauletekio al. 3, LT-10222 Vilnius, Lithuania

e-mail: viktorija.gineityte@tfai.vu.lt        Phone: +37052234656            :

**Abstract**
Individual Kekulè valence structures of biphenylene and related hydrocarbons are treated perturbatively by modelling them as sets of weakly-interacting uniform double (C=C) bonds. Total $\pi$-electron energies of these structures are then expressed in the form of power series with respect to the resonance parameter of uniform single (C–C) bonds. On this basis, the Kekulè structures concerned are ordered according to their relative stabilities and thereby importances when building up the actual electronic structures. To rationalize the results, interrelations are sought between separate members of the power series, on the one hand, and presence of definite substructures in the given Kekulè structure, on the other hand. It is shown that monocycles containing two and four exocyclic methylene groups [like 3,4-dimethylene cyclobutene and [4]radialene] participate in the formation of energy corrections of the relevant Kekulè valence structures along with the usual rings consisting of C=C and C–C bonds alternately and known as conjugated circuits. An extension of the empirical Fries rule to the case of biphenylene-like hydrocarbons is consequently formulated that embraces monocycles of the above-specified types. The reasons are also discussed why the results of the usual theory of conjugated circuits are less satisfactory for phenylenes as compared to benzenoids.

## 1. Introduction

Kekulè valence structures are among basic concepts of theoretical chemistry that are widely used in qualitative discussions of experimental phenomena, as well as in development of important generalities, such as aromaticity and local aromaticity [1-3]. These structures also played a crucial role in the history of valence bond (VB) theory [4-6]. Recent extensions and applications of the concept may be found in Refs. [7-9]. Since the majority of conjugated organic molecules are representable by several (or even numerous) distinct Kekulè structures [2,3], the central question usually is about relative importances (weights) of the latter. This especially refers to (poly)cyclic compounds including hydrocarbons [3].

Unfortunately, no source of direct information exists concerning the above-mentioned principal problem. It is then no surprise that an assumption about equal (uniform) weights of all Kekulè valence structures of a certain compound was the first and most straightforward solution. For example, early versions of the VB theory are based on this simple assumption [5] along with definitions both of the Pauling bond order [10] and of the Kekulè structure count (KSC) $K$ [4, 11]. Besides, dependences of molecular characteristics upon $K$ are still under interest, e.g. of the total $\pi$-electron energy [12-14].

An alternative option evidently consists in looking for criteria to select the structure(s) of the high(est) weight(s), even more so because "it is generally accepted that not all Kekulè valence structures are equally important" [2]. This direction of the research is primarily represented by simple rules of Fries [15, 16] and Clar [17, 18] originally suggested for benzenoid hydrocarbons and based on counting of benzenoid rings and of disjoint $\pi$ sextets, respectively, in the given Kekulè structure. Further development of the same ideas may be exemplified by the concepts of conjugated circuits [1,3] and of maximum valence structures [2]. The weights of individual Kekulè valence structures are correspondingly determined here by the relevant partial contributions to the total molecular resonance energy (RE) [3] and "by the smallest Pauling bond order found for any CC double bond" [2]. More sophisticated criteria, in turn, may be classified into algebraic and numerical ones. The former may be exemplified by the parity of a certain Kekulè structure along with the so-called algebraic structure count (ASC) [3, 19. 20], as well as by the innate degree of freedom (df) [21]. Meanwhile, the Kekulè index [22, 23], connectivity of submolecules underlying individual Kekulè structures [24-26] and energetic parameters following from *ab intio* VB calculations [27] represent numerical criteria.

A noteworthy point in the present context is that the principal problem of selecting the most (more) important Kekulè structure(s) actually consists of two parts: First, a reliable criterion is required to distinguish between structures, the extent of similarity of which is as high as possible. Second, the resulting distinctions should be explained (rationalized) in terms of specific aspects of constitution of the structures under comparison so that definite rules may be subsequently formulated (such as those of Fries and Clar). A particular approach may be then accordingly characterized by its discriminative potential and explanatory power.

Analysis of the above-overviewed criteria shows that a high discriminative potential usually goes together with a low explanatory power and vice versa. For example, no explicit relations seem to be obtainable between relative values of numerical criteria (e. g. of the Kekulè index) and constitutional parameters of the structures concerned in spite of their high discriminative potentials. Again, the simple and illustrative Fries rule is characterized by a comparatively low discriminative potential (because several structures of an extended compound often contain the same number of benzenoid rings). That is why our study is orientated towards finding an optimum combination of these two principal abilities. To this end, we invoke a new perspective on Kekulè valence structures themselves, viz. the perturbational one. Before turning to the latter, let us recall the definition of the very Kekulè structure.

Let us confine ourselves to the case of conjugated hydrocarbons for simplicity. In the VB theory, an individual Kekulè structure of these compounds represents a particular scheme of pairing of $\pi$-electrons [5, 6]. So far as qualitative structural chemistry in general is concerned, the same structure is usually understood as one of several possible arrangements of C=C and C–C bonds. At the Hűckel level, the second definition implies a certain pattern of alternating resonance parameters. Given that parameters of single bonds are supposed to take sufficiently small relative values (vs. those of double bonds) in addition, the resulting model of an individual Kekulè structure coincides with a set of weakly- interacting (perturbed) C=C bonds, where the C–C ones play the role of the interaction. Accordingly, perturbative approaches may be applied to express the relevant total $\pi$-electron energy in the form of power series with respect to the above-specified small parameters. Relative stabilities of distinct Kekulè structures and thereby their weights then follow from comparison of these energies. Besides, an analogous (i.e. perturbational) perspective on acyclic conjugated hydrocarbons (polyenes) is widely known and exploited [28-30]. This analogy is even more evident if we recall that the only Kekulè structures of isomers usually are under comparison in the latter case. Thus, experience of these studies deserves now our attention.

The simple perturbational MO (PMO) theory [28] apperently is the earliest perturbative approach to $\pi$-electron systems of acyclics. This theory, however, meets significant difficulties in distinguishing between total $\pi$-electron energies of isomers [28, 30]. In this connection, more sophisticated approaches have been suggested in this field [31-35]. These have been comparatively discussed previously [36] along with advantages of the power series derived in Refs. [33-35] without invoking the delocalized (canonical) molecular orbitals [The derivation was carried out directly in the basis of two-center orbitals of double (C=C) bonds called the bond orbitals for simplicity (Sect. 2)]. Application of just this series allowed us to discriminate between stabilities of slightly different isomers of numerous extended polyenes [36, 37], as well as to account for reduced stabilities of branched ones [36]. These achievements point to both a high discriminative potential and an outstanding explanatory power of the approach.

A similar study of individual Kekulè structures of benzenoid hydrocarbons [35] also supports the above statement. In particular, explicit relations have been revealed in this case between members of the power series for their total energies and presence of specific cyclic substructures in the given Kekulè structure. These results provided us with a perturbative analogue for the simple theory of conjugated circuits [1, 3].

To continue studies in the same direction, we now turn to biphenylene and related hydrocarbons [38]. It deserves an immediate emphasizing that this case is much more complicated as compared to that of benzenoids. Indeed, both biphenylene and its derivatives contain four-atomic rings along with six-atomic ones, appearently contributing to destabilization and to stabilization of the relevant $\pi$-electron system, respectively (provided that the classical 4n+2/4n rule [39, 40] is extendable to polycyclic systems). Thus, an interplay of opposite factors seems to determine the actual REs in this case [41, 42]. It is then no surprise that even the extent and the nature of aromaticity of the parent biphenylene are still under discussion [3, 42-44] to say nothing about its derivatives (e. g. [n] phenylenes [45]). The relevant dependences of total $\pi$-electron energies upon the KSCs ($K$) also were found to differ substantially from that of benzenoids [13, 14]. Finally, the results of application of the theory of conjugated circuits proved to be considerably less satisfactory in the case of phenylenes [3, 45]. This especially refers to the simplest version of this theory, wherein all Kekulè structures of a certain compound are initially assumed to be of the same weight. All these circumstances stimulate an interest in development of the perturbational perspective on Kekulè valence structures of these non-benzenoid molecules.

In summary, our aim consists in application of the perturbative approach of Refs. [33-35] to individual Kekulè structures of biphenylene-like hydrocarbons to order these structures according to their stabilities (and thereby importances), as well as to reveal the principal substructures (fragments) that are responsible for the relevant energetic distinctions.

## 2. Theory

Let us start with our model of Kekulè valence structures. The principal assumptions underlying the latter are as follows: i) The structures concerned consist of an even number ($2N$) of uniform (carbon) atoms, as well as contain two types of uniform bonds, namely $N$ double (C=C) bonds and $N'$ single (C–C) ones [Note that $N$ generally does not coincide with $N'$]; ii) The single (C–C) bonds are weak as compared to the double (C=C) ones; iii) The structures contain no cycles embracing an odd number of atoms and thereby belong to even alternant hydrocarbons [28, 39, 40].

At the Hűckel level, the above-enumerated assumptions yield the following implications: i) The structures under study are representable by three principal parameters in the $2N$-dimensional basis of $2p_z$ AOs of carbon atoms $\{\chi\}$, namely by the Coulomb parameter ($\alpha$), as well as by resonance parameters $\beta$ and $\gamma$ corresponding to double and single bonds, respectively. Given that the usual equalities $\alpha = 0$ and $\beta = 1$ are accepted for convenience, a negative energy unit and thereby a positive $\gamma$ value ($\gamma > 0$) also is among implications; ii) The parameter $\gamma$ takes a sufficiently small value so that it may be regarded as a first order term vs. our energy unit; iii) The basis set $\{\chi\}$ may be divided into two $N$-dimensional subsets $\{\chi^*\}$ and $\{\chi^o\}$ so that pairs of orbitals belonging to chemical bonds (C=C or C–C) find themselves in different subsets [Note that resonance parameters between pairs of AOs of chemically-bound atoms only are taken into consideration as usual [39, 40]]. This, in turn, implies the non-zero resonance parameters (i.e. both 1 and $\gamma$) to take place in the off-diagonal (intersubset) blocks of our initial Hamiltonian matrix ($H$) as exhibited below in Eq. (1). Meanwhile, the relevant diagonal (intrasubset) blocks coincide with zero matrices; iv) The AOs always may be enumerated in such a way that orbitals belonging to the same C=C bond (say, to the Ith one) acquire the coupled numbers $i$ and $N+i$. The resonance parameters of these strong bonds then take the diagonal positions in the intersubset blocks of the matrix $H$.

As a result, the initial Hamiltonian matrix of our Kekulè valence structures ($H$) consists of a sum of the zero order member ($H_{(0)}$) and of the first order one ($H_{(1)}$), including parameters of C=C and C–C bonds, respectively, viz.

$$H = H_{(0)} + H_{(1)} = \begin{vmatrix} 0 & I \\ I & 0 \end{vmatrix} + \begin{vmatrix} 0 & \gamma B \\ \gamma B^+ & 0 \end{vmatrix}, \qquad (1)$$

where the orders are defined with respect to parameter $\gamma$ and correspondingly denoted by subscripts (0) and (1). Notation $I$ here and below stands for the unit (sub)matrix and the superscript + designates the transposed (Hermitian-conjugate) matrix. Unit off-diagonal elements of the (sub)matrix $B$ ($B_{ij}$, $i \neq j$) correspond to C–C bonds, otherwise these take zero values. Meanwhile, the diagonal elements vanish (i.e. $B_{ii}=0$), because entire resonance parameters of C=C bonds are included into the zero order matrix $H_{(0)}$.

The above-introduced subsets of AOs $\{\chi^*\}$ and $\{\chi^o\}$ are characterized by a zero energy gap. Consequently, the perturbative approach of Refs. [33-35] cannot be straightforwardly applied to the matrix $H$ of Eq.(1) and a definite transformation of the basis set is required. To this end, let us introduce a new basis of bond orbitals (BOs) of C=C bonds $\{\varphi\}$. The bonding BO (BBO) of the Ith C=C bond and its antibonding counterpart (ABO) will be defined as a normalized sum and difference, respectively, of the relevant AOs $\chi_i^*$ and $\chi_{N+i}^o$. Let these BOs to be correspondingly denoted by $\varphi_{(+)i}$ and $\varphi_{(-)i}$, where the subscript $i$ refers to the Ith bond. Finally, the subset of BBOs and that of ABOs will be accordingly designated by $\{\varphi_{(+)}\}$ and $\{\varphi_{(-)}\}$, respectively. Passing from the initial basis $\{\chi\}$ to the new one $\{\varphi\}$ is then carried out by means of a simple unitary matrix

$$U = \frac{1}{\sqrt{2}} \begin{vmatrix} I & I \\ I & -I \end{vmatrix}. \qquad (2)$$

Consequently, the $N$x$N$-dimensional submatrices of the transformed Hamiltonian matrix $H'$ refer to individual subsets $\{\varphi_{(+)}\}$ and $\{\varphi_{(-)}\}$ and to their interaction. Again, the new matrix $H'$ also consists of a zero order member ($H'_{(0)}$) and of a first order one ($H'_{(1)}$) with respect to the same parameter $\gamma$, viz.

$$H' = H'_{(0)} + H'_{(1)} = \begin{vmatrix} I & 0 \\ 0 & -I \end{vmatrix} + \begin{vmatrix} V & T \\ T^+ & W \end{vmatrix}, \qquad (3)$$

where

$$V = -W = \frac{\gamma}{2}(B + B^+), \qquad T = \frac{\gamma}{2}(B^+ - B). \qquad (4)$$

The zero order term of Eq.(3) ($H'_{(0)}$) contains one-electron energies of BBOs and of ABOs in its diagonal positions [these correspondingly coincide with 1 and –1 in our energy units], whereas the first order one ($H'_{(1)}$) embraces interactions between BOs of C=C bonds (interbond resonance parameters). In particular, elements of submatrices $V$, $W$ and $T$ represent the following types of the above-mentioned interactions

$$V_{ij} = \langle \varphi_{(+)i} | \hat{H} | \varphi_{(+)j} \rangle, \qquad W_{lm} = \langle \varphi_{(-)l} | \hat{H} | \varphi_{(-)m} \rangle, \qquad T_{il} = \langle \varphi_{(+)i} | \hat{H} | \varphi_{(-)l} \rangle, \qquad (5)$$

where the BOs concerned are shown inside the bra- and ket-vectors. Proportionality of these elements to $\gamma$ seen from Eq.(4) reflects the fact that the interaction between C=C bonds takes place through C–C bonds. Finally, the equality $B_{ii}=0$ along with Eq.(4) yields zero values for intrabond resonance parameters $V_{ii}$, $W_{ii}$ and $T_{ii}$ as exhibited below in Eq.(8).

As is seen from Eq.(3), the subsets $\{\varphi_{(+)}\}$ and $\{\varphi_{(-)}\}$ are now separated one from another by a substantial energy gap (equal to 2) vs. the intersubset interaction $T$. Thus, the matrix $H'$ meets the requirements of the perturbative expansion of Refs. [33-35]. Application of the latter then allows the total $\pi$-electron energy of any Kekulè valence structure ($E$) to be represented in the form of power series, i.e. as a sum of increments ($E_{(k)}$) of increasing orders $k$ with respect to the resonance parameter of C–C bonds ($\gamma$). The zero order member of this series ($E_{(0)}$) coincides with the total energy of $N$ isolated C=C bonds (equal to $2N$) in accordance with the expectation, whereas the first order one ($E_{(1)}$) vanishes [33]. An additional stabilization (or destabilization) of the given structure due to presence of C–C bonds may be then conveniently described by the difference ($\Delta E$) between the relevant total energy ($E$) and the respective zero order increment ($E_{(0)}$). The power series concerned is then as follows

$$\Delta E = \sum_{k=2}^{\infty} E_{(k)} = E_{(2)} + E_{(3)} + E_{(4)} + ... \tag{6}$$

As demonstrated in Refs. [33-35], several distinct representations are possible for members of this series. To exhibit the most important ones, let us introduce new matrices $G_{(k)}$, $k=1,2,3...$ that are expressible via those of lower orders ($G_{(k-1)}$, $G_{(k-2)}$, etc.), as well as in terms of the former matrices $V$, $W$ and $T$, viz.

$$G_{(1)} = -\frac{1}{2}T, \qquad G_{(2)} = -\frac{1}{2}(VG_{(1)} - G_{(1)}W) = \frac{1}{4}(VT - TW) = \frac{1}{4}(VT + TV),$$
$$G_{(3)} = -\frac{1}{2}(VG_{(2)} - G_{(2)}W) - G_{(1)}G_{(1)}^+ G_{(1)} = -\frac{1}{8}(V^2T + 2VTV + TV^2 + T^3), \tag{7}$$

where the relations $V=-W$ and $T^+=-T$ seen from Eq.(4) also are invoked. [Note that the notation $\tilde{G}_{(3)}$ stood fot the present matrix $G_{(3)}$ in Ref.[35]]. Substituting Eq.(4) into Eq.(7) shows that matrices $G_{(k)}$ correspondingly contain factors $\gamma^k$ and thereby these are terms of the $k$th orders as indicated by the subscripts ($k$). Zero diagonal elements of the same matrices easily follow from their skew-Hermitian nature [46]. The overall result concerning diagonal elements of our matrices takes then the form

$$V_{ii} = W_{ii} = T_{ii} = G_{(k)ii} = 0, \qquad k = 1,2,... \tag{8}$$

In terms of matrices $G_{(k)}$, the energy increments $E_{(k)}$ of Eq.(6) take the most compact form [35], viz.

$$E_{(k)} = 4Tr(G_{(k-1)}G_{(1)}^+) \equiv 4\sum_{(+)i}\sum_{(-)l} G_{(k-1)il}G_{(1)li}^+ \equiv 4\sum_{(+)i}\sum_{(-)l} G_{(k-1)il}G_{(1)il}, \tag{9}$$

where the notation $Tr$ stands for a trace of the matrix product within parentheses. The right-hand side of Eq.(9) contains elements of matrices $G_{(1)}$ and $G_{(k-1)}$, each of them connecting a BBO ($\varphi_{(+)i}$) and an ABO ($\varphi_{(-)l}$) and expressible as follows

$$G_{(1)il} = -\frac{1}{2}T_{il}, \qquad G_{(2)il} = \frac{1}{4}\left[\sum_{(+)j} V_{ij}T_{jl} - \sum_{(-)m} T_{im}W_{ml}\right], \quad etc. \tag{10}$$

Accordingly, sums of both Eq.(9) and Eq.(10) embrace either all BBOs or all ABOs of the given system.

To interpret the above expressions, let us start with elements $G_{(k)il}$. It is seen that the first order element $G_{(1)il}$ is proportional to the resonance parameter ($T_{il}$) between the relevant BOs ($\varphi_{(+)i}$ and $\varphi_{(-)l}$) and inversely proportional to the energy gap between BBOs and ABOs (equal to 2). Thus, this element represents the direct (through-space) interaction between BOs $\varphi_{(+)i}$ and $\varphi_{(-)l}$. Besides, zero values of these interactions inside the same C=C bond result from Eq.(8). Similarly, the element $G_{(2)il}$ of the matrix $G_{(2)}$ is interpretable as the indirect (through-bond) interaction of the same BOs ($\varphi_{(+)i}$ and $\varphi_{(-)l}$), wherein BBOs ($\varphi_{(+)j}$) and ABOs ($\varphi_{(-)m}$) of other bonds play the role of mediators [Note the $j \neq i$ and $m \neq l$ because of Eq.(8)]. As is seen from Eq.(10), the orbitals $\varphi_{(+)j}$ and $\varphi_{(-)m}$ should overlap directly both with $\varphi_{(+)i}$ and with $\varphi_{(-)l}$ to be efficient mediators and thereby to ensure a non-zero value of $G_{(2)il}$. Analogously, the third order element $G_{(3)il}$ represents the indirect interaction of the same BOs by means of two mediators. Generally, the total number of mediators of any element of the $k$th order coincides with $k-1$.

The above-discussed interpretation of elements $G_{(k-1)il}$ and $G_{(1)il}$ within Eq.(9) shows that the energy correction $E_{(k)}$ depends upon products of direct and indirect interactions between pairs of BOs. We then arrive at the following rule [47]:

**Rule 1**: The correction $E_{(k)}$ takes a non-zero value, if there is at least a single pair of orbitals $\varphi_{(+)i}$ and $\varphi_{(-)l}$ that interact both directly and indirectly by means of $k-2$ mediators. Moreover, the pair of BOs concerned contributes to stabilization (destabilization) of the system, if the above-mentioned interactions are of the same (opposite) signs [The negative sign of our energy unit should be recalled here].

As already mentioned, the expression of Eq.(9) is not the only form of energy corrections $E_{(k)}$. If we invoke, for example, the evident one-to-one correspondence between BBOs $\varphi_{(+)i}$ and C=C bonds, the corrections $E_{(k)}$ become alternatively representable as sums of increments ($E_{(k)I}$) of individual double bonds ($I=1,2,3…N$), i.e.

$$E_{(k)} = \sum_I E_{(k)I}, \qquad (11)$$

where

$$E_{(k)I} = 4 \sum_{(-)l \neq i} G_{(k-1)il} G_{(1)il}. \qquad (12)$$

The expression of Eq.(11) is referred to below as the additive form of $E_{(k)}$.

Finally, explicit formulae are easily obtainable for increments $E_{(k)}$ in terms of interbond resonance parameters of Eq.(5). To this end, expressions for individual elements $G_{(k-1)il}$ and $G_{(1)il}$ like those of Eq.(10) should be substituted into either Eq.(9) or Eqs. (11) and (12). Because of a rather cumbersome nature of the result [48], however, we confine ourselves here to general properties of the latter based on somewhat formalized perspective on elements $G_{(k)il}$.

As is seen from Eq.(10), the second order element $G_{(2)il}$ contains products of pairs of connected resonance parameters of Eq.(5), the third order one ($G_{(3)il}$) embraces analogous triplets of elements of matrices ***T, V*** and ***W***, and so forth. Consequently, the element $G_{(k)il}$ takes a non-zero value, if there is at least a single non-zero product of resonance parameters of the following form

$$\langle \varphi_{(+)i} | \hat{H} | \varphi_1 \rangle \langle \varphi_1 | \hat{H} | \varphi_2 \rangle \langle \varphi_2 | \hat{H} | \varphi_3 \rangle .... \langle \varphi_{k-2} | \hat{H} | \varphi_{k-1} \rangle \langle \varphi_{k-1} | \hat{H} | \varphi_{(-)l} \rangle \neq 0, \qquad (13)$$

where $\varphi_1, \varphi_2,...\varphi_{k-1}$ stand for mediating orbitals. Given that the condition of Eq.(13) is met, we will say that in the given system there is a pathway of the $k$th order between BOs $\varphi_{(+)i}$ and $\varphi_{(-)l}$. It also deserves emphasizing that steps inside the same C=C bond are not allowed in this pathway owing to Eq.(8). Additivity of elements $G_{(k)il}$ with respect to different pathways (if any) also is noteworthy. Further, employment of these concepts within Eq.(12) shows that the contribution of the Ith C=C bond ($E_{(k)I}$) to the energy correction $E_{(k)}$ is determined by products of pathways of the first and of the ($k$–1)th orders embracing the BBO of this double bond ($\varphi_{(+)i}$) and ABOs of the remaining bonds ($\varphi_{(-)l}$). Thus, we may define a self-returning pathway of the $k$th order both starting and terminating at the BBO $\varphi_{(+)i}$ and embracing the ABO $\varphi_{(-)l}$ that is able to replace the above-mentioned product of two linear pathways. Our principal statement is then as follows:

**Rule 2**: The contribution of the Ith bond ($E_{(k)I}$) to the $k$th order energy ($E_{(k)}$) takes a non-zero value if there is at least a single bond (say, the Lth one), the ABO of which participates in a self-returning pathway of the $k$th order both starting and terminating at the BBO $\varphi_{(+)i}$. Moreover, both $E_{(k)I}$ and $E_{(k)}$ are additive quantities with respect to increments of individual self-returning pathways over BOs. [Besides, rules governing the signs of increments of particular self-returning pathways also may be formulated in some cases by extending the concepts of Hückel and Möbius aromaticity to cycles consisting of BBOs and ABOs [48]].

Let us recall now that $k$ basis functions (BOs) participate in the formation of an element $G_{(k-1)il}$, viz. $\varphi_{(+)i}$, $\varphi_{(-)l}$ and $k$–2 madiators. The same then refers to the energy increment $E_{(k)I}$ of Eq.(12). Given that all these BOs belong to distinct C=C bonds, the increment $E_{(k)I}$ embraces exactly $k$ C=C bonds. Otherwise, this number is lower than $k$. Our last rule is then as follows:

**Rule 3**: Substructures (fragments) consisting of no more than $k$ C=C bonds participate in the formation of the total $k$th order energy $E_{(k)}$. Given that convergence of the power series of Eq.(6) [46] is taken into consideration in addition, extinction of increments of individual substructures is foreseen when the size of the latter grows.

For illustration, let us consider the starting members of the series of Eq.(6). The second order term $E_{(2)}$ is determined by squares of first order elements $G_{(1)il}$ as Eq.(9) indicates. Thus, this term is an *a priori* positive quantity contributing to stabilization of the given system. Since each element ($G_{(1)il}$) embraces two BOs ($\varphi_{(+)i}$ and $\varphi_{(-)l}$), the second order energy $E_{(2)}$ contains a sum of contributions ($E_{(2)IL}$) of individual pairs of double (C=C) bonds (I and L). Again, a simple self-returning pathway from the BBO $\varphi_{(+)i}$ to the ABO $\varphi_{(-)l}$ and backwards corresponds to any term $(G_{(1)il})^2$ and/or $G_{(1)il} G_{(1)li}^+$. The correction $E_{(2)}$ is then accordingly expressible as a sum of increments of these pathways. Our previous study of Kekulè valence structures of both benzenoids [35] and polyenes [36] showed that non-zero increments to $E_{(2)}$ actually originate only from pairs of the so-called first-neighboring C=C bonds coinciding with those connected by a C–C bond. Moreover, these significant increments are uniform and equal to $\gamma^2/2$, the latter value coinciding with the second order energy of the only Kekulè structure of the linear butadiene [34, 35]. As a result, the correction $E_{(2)}$ of the above-specified extended Kekulè structures proved to be proportional to the relevant total numbers of C–C bonds ($N'$), viz.

$$E_{(2)}^{ac} = \frac{1}{2}\gamma^2 N', \qquad (14)$$

where the meaning of the superscript *ac* is clarified below (Subsect. 3.2).

Analogously, the third order term $E_{(3)}$ is determined by products $G_{(2)il}G_{(1)li}^+$ and/or $G_{(2)il}G_{(1)il}$ as Eq.(9) shows. This implies this member of the series to take a non-zero value if in the given system there is at least a single pair of BOs ($\varphi_{(+)i}$ and $\varphi_{(-)l}$) that interact both directly and indirectly by means of a single mediator (see the Rule 1). Moreover, a particular product $G_{(2)il}G_{(1)il}$ yields a positive (negative) contribution to the total correction $E_{(3)}$, if the participating interbond interactions (i.e. both $G_{(2)il}$ and $G_{(1)il}$) are of the same (opposite) signs. Since the mediating orbital necessarily belongs to a third (say, Mth) C=C bond (that coincides neither with the Ith bond nor with the Lth one), an alternative form of the same non-zero-value condition for $E_{(3)}$ resolves itself into a requirement of presence of at least a single triplet of C=C bonds (I, L, M), the BOs of all pairs of which interact (overlap) directly and/or of at least a single self-returning pathway over BOs of the third order. For example, pathways of the following types

$$\langle \varphi_{(+)i}|\hat{H}|\varphi_{(+)m}\rangle\langle \varphi_{(+)m}|\hat{H}|\varphi_{(-)l}\rangle\langle \varphi_{(-)l}|\hat{H}|\varphi_{(+)i}\rangle \neq 0,$$
$$\langle \varphi_{(+)i}|\hat{H}|\varphi_{(-)m}\rangle\langle \varphi_{(-)m}|\hat{H}|\varphi_{(-)l}\rangle\langle \varphi_{(-)l}|\hat{H}|\varphi_{(+)i}\rangle \neq 0 \qquad (15)$$

embrace the above-specified triplet of C=C bonds (I, L, M) and thereby contribute to the increment $E_{(3)I}$ of the third order energy $E_{(3)}$.

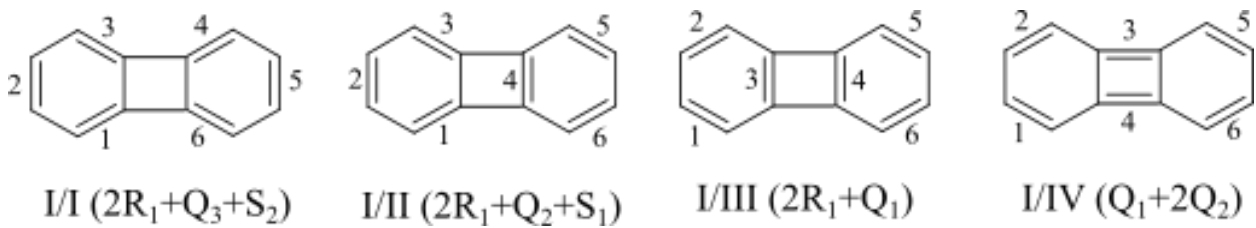

Figure 1. Symmetry-non-equivalent Kekulè valence structures of biphenylene (I). Numbers of C=C bonds also are shown along with monocycles contained.

## 3. Results and discussion

The initial part of our discussion is devoted to comparison of total $\pi$-electron energies of Kekulè structures of biphenylene-like hydrocarbons, i.e. to the discriminative aspect of the problem (Sect. 1).

### 3.1. Ordering of Kekulè valence structures

Let us start with biphenylene itself (I). The four symmetry-non-equivalent Kekulè structures of this hydrocarbon (I/I, I/II, I/III and I/IV) are shown in Fig. 1, all of them containing six C=C bonds ($N$=6) and eight C–C bonds ($N'$=8). [It deserves attention that $N \neq N'$]. As a result, the relevant zero order energies also are uniform and coincide with 12 in our negative energy units. Thus, we may turn immediately to comparison of stabilization energies $\Delta E$ defined by Eq.(6). Separate terms of this series for structures I/I-I/IV are shown in Table 1 along with the total energy differences ($\Delta E$) for reasonable values of our parameter $\gamma$, viz. for $\gamma = 0.1$ and $\gamma = 0.2$. It is seen that the first two structures (I/I and I/II) are predicted to be more stable as compared to the remaining ones (I/III and I/IV) on the basis of the second order energies $E_{(2)}$. After taking into account the third order corrections ($E_{(3)}$) inside the above-specified couples, we obtain the following order of stability: I/I>I/II>I/III>I/IV. Thus, sums of second and third order energies actually are sufficient to discriminate between the structures concerned. Moreover, the above-concluded order is entirely supported by numerical $\Delta E$ values, wherein the fourth order corrections ($E_{(4)}$) also are incorporated. Independence of the result upon the particular choice of the parameter $\gamma$ also is noteworthy. Most importantly, our order of stability of structures I/I-I/IV coincides with those following from other approaches, including theories of conjugated circuits [3] and of the maximum valence structure [2], as well as from algebraic and numerical criteria, such as parity [19], innate degrees of freedom (df) [21] and Kekulè index [22, 23]. Finally, the same order has been established recently on the basis of a modified Hess-Schaad group additivity scheme [42].

Kekulè valence structures of related hydrocarbons also may be ordered analogously. In particular, the case of naphthocyclobutene (II) (Fig. 2) closely resembles that of biphenylene (I) in respect of sufficiency of sums of second and third order terms. The resulting order of stability (viz. II/I>II/II>II/III>II/IV) coincides with that following from Kekulè indices [22]. Moreover, uniform zero order energies of all structures of both molecules (i.e. of I and II) allows direct comparisons of their stabilities, e.g. the structures I/II and II/I prove to be isoenergetic to within fourth order terms inclusive.

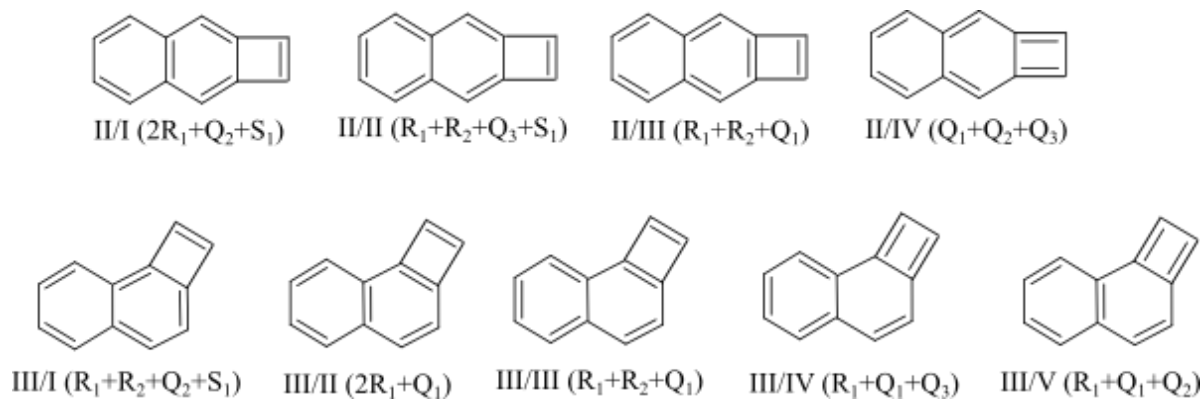

Figure 2. Kekulè valence structures of naphthocyclobutenes II and III along with the monocycles present there.

A somewhat different state of things is in the case of the bent isomer of naphthocyclobutene (III) characterized by five distinct Kekulè structures III/I-III/V (Fig. 2). As is seen from Table 1, sums $E_{(2)}+E_{(3)}$ are uniform for structures III/III, III/IV and III/V and, consequently, the subsequent terms of the power series start to play the decisive role. Fortunately, the relevant fourth order corrections take different values so that the final order of stability is as follows: III/I>III/II>III/III>III/IV>III/V. An analogous conclusion has been drawn also on the basis of Kekulè indices [22] and connectivities of the relevant submolecules [26].

More extended phenylenes are exemplified by bent [3]phenylene (IV), four selected Kekulè structures of which are shown in Fig. 3. The respective sums $E_{(2)}+E_{(3)}$ (Table 1) suggest that the structures IV/I and IV/IV take the first and the last positions, respectively, in the relevant order of stability. To discriminate between energies of the remaining structures (IV/II and IV/III), however, fourth order corrections should be invoked. The result is then as follows: IV/I>IV/II>IV/III>IV/IV. This outcome is in line with actual contributions of the structures to the molecular RE [2,3].
Finally, the three Kekulè valence structures of benzocyclobutene (V) (Fig. 4) also are included into our collection. The resulting order of stability (V/I>V/II>V/III) is in line with that following from the theory of conjugated circuits [3]. These relatively simple structures serve as model systems in our further discussion (Subsect. 3.3).

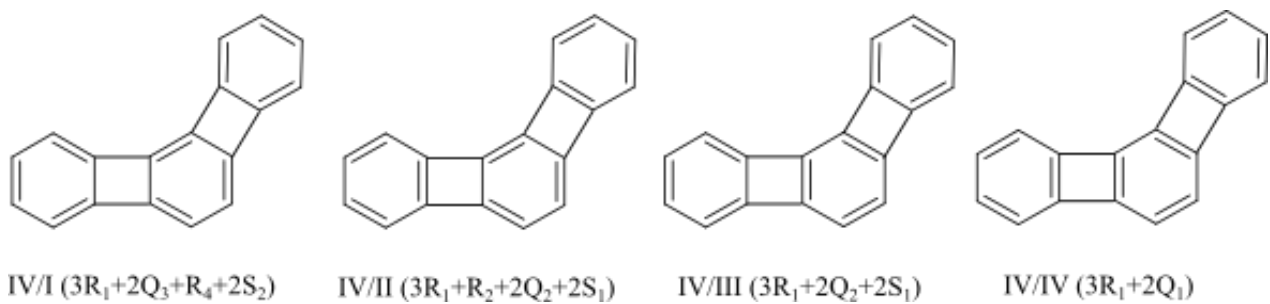

Figure 3. Selected Kekulè valence structures of bent [3]phenylene (IV) and their compositions in terms of monocycles.

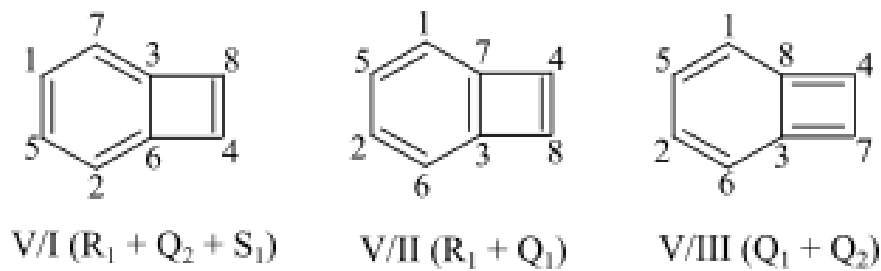

Figure 4. Kekulè valence structures of benzocyclobutene (V). Numbers of carbon atoms and/or of their 2p$_z$ AOs also are shown along with the relevant monocycles.

On the whole, the above results demonstrate adequacy of the perturbative approach applied in evaluations of relative stabilities of Kekulè valence structures. Moreover, a high discriminative potential of the approach is evident. It is also seen that lowest stabilities and/or importances are predicted for Kekulè structures in which either the neighboring hexagonal rings are connected by two C=C bonds (such as I/IV) or two exocyclic C=C bonds are attached to a certain hexagonal ring (e.g. II/IV, III/V, etc.). Structures of just this type have been suggested to be excluded when applying the theory of conjugated circuits to extended phenylenes [45]. Moreover, a low importance of these structures follows also from the Clar rule [17, 18]. Another common property of the above results is that total energies of individual Kekulè valence structures of biphenylene-like hydrocarbons differ one from another in terms of the second order. Meanwhile, the analogous distinctions were shown to be of the third order magnitude in the case of benzenoids [35]. This implies that an assumption about uniform weights of all Kekulè valence structures is less justified for biphenylene-like hydrocarbons as compared to benzenoids. Such a conclusion, in turn, serves as a deductive accounting for less satisfactory results of the simplest version of the theory of conjugated circuits for non-benzenoids (Sect. 1).

Our next aim consists in rationalization of the above-established orders of stability of Kekulè valence structures. To this end, we will look for relations between separate energy increments ($E_{(k)}$), on the one hand, and constitution of the given Kekulè structure, on the other hand. Since the constitution concerned is representable by presence of specific monocyclic substructures as experience of related approaches shows [c. f. the benzenoid rings underlying the rule of Fries [15,16]], we will turn now to an overview of energy increments ($E_{(k)}$) referring to isolated monocycles contained in the Kekulè structures of hydrocarbons I-V and exemplified in Fig.5.

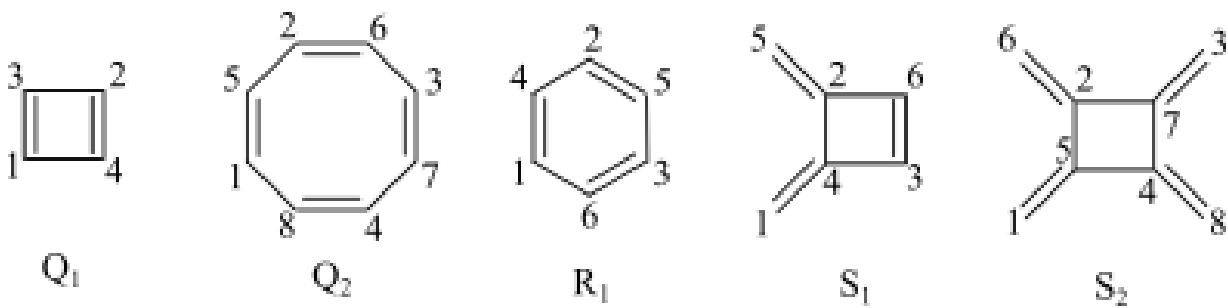

Figure 5. Examples of monocyclic substructures contained in the Kekulè valence structures of hydrocarbons I-V. Numbers of carbon atoms and/or of their 2p$_z$ AOs also are shown.

### 3.2. Energy increments for isolated monocyclic substructures

Let us start with general properties of pathways over BOs of monocycles. First, two types of self-returning pathways reveal themselves in this case, viz. the cyclic (roundabout) and the acyclic ones [48, 49]. Let the former to be defined as a pathway which embraces the whole cycle in a clockwise or anticlockwise fashion so that each C=C bond is visited only once. Otherwise, we have to do with a pathway of "toward-backward" nature, which embraces a certain linear (acyclic) fragment of the cycle. Second, cyclic (roundabout) pathways are of the $mN$-th orders in a monocycle containing $N$ double (C=C) bonds, where $m$=1,2,… is an integer number. The shortest pathway of this type accordingly is of the $N$th order [Note that steps inside the same C=C bond are not allowed (Sect. 2)]. Third, self-returning acyclic pathways of even orders only are possible in the systems concerned.

Additivity of terms of power series for total energies with respect to increments of individual pathways (Rule 2) then consequently permits each correction ($E_{(k)}$) to be generally represented as a sum of two components, viz.

$$E_{(k)} = E_{(k)}^{ac} + E_{(k)}^{cycl},\qquad(16)$$

where the superscripts *ac* and *cycl* correspondingly indicate their acyclic and cyclic origin. The component $E_{(k)}^{cycl}$ is also called below the cyclization energy of the given cycle. Moreover, the actual values of these components are governed by the following selection rules: (i) The acyclic components ($E_{(k)}^{ac}$) vanish for odd values of the order parameter $k$; (ii) The cyclic components ($E_{(k)}^{cycl}$) take non-zero values for $k=mN$ only. Since the increments $E_{(2N)}^{cycl}$, $E_{(3N)}^{cycl}$, etc. are of little interest because of their relatively high orders, an $N$-membered monocycle proves to be primarily characterized by the cyclization energy of the $N$th order ($E_{(N)}^{cycl}$).

Members of power series for total $\pi$-electron energies of individual monocycles present in our Kekulè valence structures (Table 2) are in line with the above-formulated rules. To show this, let us consider cycles of particular types separately. Let us start with the usual rings consisting of C=C and C–C bonds alternately. Two types of these monocycles may be subsequently distinguished on the basis of parity of their principal parameter $N$, namely the odd- and even-membered ones [Note that $N = N'$ in this case]. Since the monocycles concerned correspondingly coincide with conjugated circuits of the 4n+2 and 4n series, the commonly-accepted designations of the latter [1,3] will be used, viz. $R_n$ and $Q_n$, n=1,2,3… [4n+2 and 4n stand here for the relevant numbers of $\pi$-electrons and/or of carbon atoms].

Let us dwell first on the odd-membered rings $R_n$. The first representative of this series $R_1$ (Fig. 5) coincides with a single Kekulè valence structure of benzene (benzenoid ring), the second one $R_2$ embraces five C=C bonds, etc. Roundabout pathways over BOs also are of odd orders in this case. Consequently, the relevant second and fourth order energies are of acyclic origin. Moreover, additivity of these corrections with respect to contributions of individual C=C bonds (see Eq. (11)) along with transferability of the latter due to regular constitution of our rings $R_n$ give birth to the following relations [35]

$$E_{(2)}(R_n) = E_{(2)}^{ac}(R_n) = \frac{1}{2}N\gamma^2, \quad E_{(4)}(R_n) = E_{(4)}^{ac}(R_n) = \frac{1}{32}N\gamma^4, \quad (17)$$

where $\gamma^2/2$ and $\gamma^4/32$ are the relevant increments of a double bond. Besides, the equality $N = N'$ ensures equivalence between the first relation of Eq.(17) and that of Eq.(14), both indicating proportionality of the correction concerned to the size of the cycle seen from the relevant part of Table 2. By contrast, most of the third order energies of the same cycles vanish, except for $E_{(3)}(R_1)$ coinciding with the cyclization energy $E_{(3)}^{cycl}(R_1)$ of the three-membered ring $R_1$ and equal to $3\gamma^3/4$. Analysis of the explicit expression for this exclusive correction shows that 12 roundabout pathways like those of Eq.(15) participate in its formation, all of them contributing to stabilization of the ring. The stabilizing nature of the correction $E_{(3)}(R_1)$ may be also alternatively traced back to proportionality [35] between the underlying matrices (see Eq.(9)), viz.

$$\boldsymbol{G}_{(2)}(R_1) = \frac{1}{2}\gamma\boldsymbol{G}_{(1)}(R_1), \quad (18)$$

where

$$\boldsymbol{G}_{(1)}(R_1) = -\frac{\gamma}{4}\begin{vmatrix} 0 & 1 & -1 \\ -1 & 0 & 1 \\ 1 & -1 & 0 \end{vmatrix}. \quad (19)$$

Due to Eq.(18), elements $G_{(1)il}$ and $G_{(2)il}$ are of uniform signs for any pair of BOs $\varphi_{(+)i}$ and $\varphi_{(-)l}$ (Rule 1) and a significant positive correction $E_{(3)}(R_1)$ results. This example demonstrates a kind of complementarity beween alternative representations of energy corrections (Sect. 2).

Let us turn now to the even-membered rings $Q_n$, n=1,2,.... Roundabout pathways over BOs also are of even orders in this case and thereby these are able to contribute to energy corrections $E_{(2)}(Q_n)$ and $E_{(4)}(Q_n)$ along with the relevant acyclic pathways. As a result, these corrections actually contain two non-zero components of Eq.(16) as discussed below. Meanwhile, the increments $E_{(k)}(Q_n)$ of odd orders ($k$) vanish for any n because of absence of appropriate pathways over BOs.

The simplest two-membered ring $Q_1$ (Fig. 5) coincides with a Kekulè valence structure of cyclobutene and contains roundabout pathways of the second order. Consequently, the relevant energy increment ($E_{(2)}(Q_1)$) consists of the following components (see Ref. [50] for details)

$$E_{(2)}^{cycl}(Q_1) = -\gamma^2, \quad E_{(2)}^{ac}(Q_1) = \gamma^2. \quad (20)$$

The negative sign and thereby the destabilizing nature of the cyclization energy $E_{(2)}^{cycl}(Q_1)$ causes little surprise if we recall the classical 4n+2/4n rule [39, 40]. Meanwhile, the acyclic component $E_{(2)}^{ac}(Q_1)$ is a positive quantity proportional to the size of the cycle ($N = N' = 2$) in accordance with the expectation (see Eq.(14)). Further, the components of Eq.(20) are of the same absolute value and cancel out one another in the final expression for the total second order energy $E_{(2)}(Q_1)$. The latter then takes an exceptional (viz. zero) value (Table 2) in spite of two C–C bonds present in the ring $Q_1$. Destabilization of the cycle $Q_1$ vs. the linear butadiene ($E_{(2)} = \gamma^2/2$ [34, 35]) also is evident. Besides, the principal matrices of the ring concerned vanish, viz.

$$\boldsymbol{G}_{(1)}(Q_1) = \boldsymbol{G}_{(2)}(Q_1) = \boldsymbol{G}_{(3)}(Q_1) = \boldsymbol{0} . \tag{21}$$

The left relation of Eq.(21) points to cancellation of direct interactions of BOs of C=C bonds via the two C–C bonds, whilst the right one ensures a zero value of the fourth order energy $E_{(4)}(Q_1)$ (see Eq.(9)).

Analogously, the four-membered cycle $Q_2$ (Fig. 5) contains roundabout pathways over BOs of the fourth order. The relevant second order energy $E_{(2)}(Q_2)$ is then of acyclic origin and proves to be proportional to the size of the cycle ($N = N' = 4$), whereas the fourth order correction $E_{(4)}(Q_2)$ contains two components, viz.

$$E_{(4)}^{cycl}(Q_2) = -\frac{20}{32}\gamma^4, \qquad E_{(4)}^{ac}(Q_2) = \frac{4}{32}\gamma^4 . \tag{22}$$

Thus, the negative cyclization energy predominates over the positive cycle-size-dependent acyclic component in this case and the total correction $E_{(4)}(Q_2)$ consequently is of destabilizing nature. The negative sign of $E_{(4)}(Q_2)$ was also shown [50] to follow directly from the vanishing second order matrix $\boldsymbol{G}_{(2)}(Q_2)$ due to cancellation of indirect interactions of BOs of second-neighboring C=C bonds (e.g. of $C_1=C_5$ and $C_3=C_7$) via orbitals of two intervening bonds ($C_2=C_6$ and $C_4=C_8$, respectively) [The point is that a zero matrix $\boldsymbol{G}_{(2)}$ always is accompanied by a negative fourth order term $E_{(4)}$ [34, 36, 37]].

Finally, the corrections of Table 2 referring to the six-membered ring $Q_3$ (i.e. both $E_{(2)}(Q_3)$ and $E_{(4)}(Q_3)$) are of acyclic and cycle-size-dependent nature in accordance with the expectation.

Apart from the above-overviewed usual rings $R_n$ and $Q_n$, monocycles containing two and four exocyclic C=C bonds also are present in the Kekulè structures under study, e.g. in the structures I/II and I/I of biphenylene (I), respectively. Let these cycles to be correspondingly denoted by $S_1$ and $S_2$ (Fig. 5). Further, an analogy is evident between substructures $S_1$ and $R_1$, as well as between $S_2$ and $Q_2$ in respect of both total numbers of C=C (and C–C) bonds and cyclic overall arrangement of the latter. This implies coinciding orders of roundabout pathways inside these couples that equal to $k=3$ and $k=4$, respectively. As a result, the relevant energy increments also are of similar nature.

Let us dwell first on the three-membered cycles $S_1$ and $R_1$, both of them characterized by cyclization energies of the third order (Table 2). The absolute value of $E_{(3)}^{cycl}(S_1)$, hovewer, is three times lower as compared to that of $E_{(3)}^{cycl}(R_1)$. This ratio may be unambiguously traced back to cancellation of contributions of some roundabout pathways over BOs in the case of $S_1$. Further, the principal matrices of the new cycle are as follows

$$\boldsymbol{G}_{(1)}(S_1) = -\frac{\gamma}{4}\begin{vmatrix} 0 & 1 & 1 \\ -1 & 0 & -1 \\ -1 & 1 & 0 \end{vmatrix}, \quad \boldsymbol{G}_{(2)}(S_1) = -\frac{\gamma^2}{8}\begin{vmatrix} 0 & -1 & 0 \\ 1 & 0 & 0 \\ 0 & 0 & 0 \end{vmatrix} \tag{23}$$

and deserve comparison to $\boldsymbol{G}_{(1)}(R_1)$ and $\boldsymbol{G}_{(2)}(R_1)$, respectively, shown in Eqs.(18) and (19). It is seen that first order matrices $\boldsymbol{G}_{(1)}(S_1)$ and $\boldsymbol{G}_{(1)}(R_1)$ are similar except for signs of some elements. Thus, BBOs and ABOs of all pairs of C=C bonds interact directly in both substructures and this fact ensures coincidence of second order energies $E_{(2)}(S_1)$ and $E_{(2)}(R_1)$ (Table 2). Meanwhile, the matrix $\boldsymbol{G}_{(2)}(S_1)$ contains only two non-zero elements ($G_{(2)12}(S_1)$ and $G_{(2)21}(S_1)$) instead of six ones of its counterpart $\boldsymbol{G}_{(2)}(R_1)$. This fact serves as an alternative accounting for the reduced absolute value of $E_{(3)}^{cycl}(S_1)$ vs. $E_{(3)}^{cycl}(R_1)$. Accordingly, the negative sign of $E_{(3)}^{cycl}(S_1)$ is due to opposite signs of matrix elements $G_{(2)12}(S_1)$ and $G_{(1)12}(S_1)$, as well as of $G_{(2)21}(S_1)$ and $G_{(1)21}(S_1)$. Since the decisive indirect interactions $G_{(2)12}(S_1)$ and $G_{(2)21}(S_1)$ are mediated here by BOs of the endocyclic ($C_3=C_6$) bond, the destabilizing nature of $E_{(3)}^{cycl}(S_1)$ may be also ascribed to the negative mediating effect of BOs of this particular bond. By contrast, the former cycle $R_1$ offers us an example of the positive mediating effect of BOs of the remaining bond (e.g. of $C_3=C_6$) in the indirect interactions between those of any pair of first-neighboring C=C bonds (i.e. of $C_1=C_4$ and $C_2=C_5$). [Distinction between a positive mediating effect and a negative one proves to be especially fruitful if the structure concerned contains a combination of both cycles $S_1$ and $R_1$ (Subsect. 3.3)]. A detailed comparative analysis of the fourth order energies $E_{(4)}(S_1)$ and $E_{(4)}(R_1)$ may be found in Ref.[50].

By contrast, the four-membered cycles $S_2$ and $Q_2$ are isoenergetic to within fourth order terms inclusive and characterized by vanishing second order matrices, viz.

$$\boldsymbol{G}_{(2)}(S_2) = \boldsymbol{G}_{(2)}(Q_2) = \boldsymbol{0} . \tag{24}$$

With the above-overviewed regularities in mind, let us now return again to Kekulè structures of biphenylene and related hydrocarbons.

## 3.3. Interpretation of results concerning biphenylene-like hydrocarbons

The actual sets of monocycles contained in separate Kekulè valence structures of hydrocarbons I-V are shown in Figs.1-4. Let us now discuss energy increments of Table 1 representing these structures in terms of contributions of individual monocycles present there.

Let us start with second order energies $E_{(2)}$. Two types of Kekulè structures reveal themselves in respect of relative values of these energies, viz. (i) the $Q_1$-free structures (e.g. I/I, I/II, II/I, etc.) characterized by corrections $E_{(2)}$ following from Eq.(14) for the relevant numbers of C–C bonds ($N'$) and (ii) the $Q_1$-containing structures (e.g. I/III, I/IV, etc.) that are destabilized vs. the former. Moreover, the extent of this destabilization is proportional both to the cyclization energy of an isolated cycle $Q_1$ ($E_{(2)}^{cycl}(Q_1)$ of Eq.(20)) and to the total number of these cycles in the structure concerned denoted below by $n(Q_1)$ [For example, the $2Q_1$-containing structure IV/IV is destabilized by $2\gamma^2$ vs. the remaining ones (i.e. IV/I-IV/III)]. In summary, the second order energies of Table 1 meet the following relation

$$E_{(2)} = E_{(2)}^{ac} + E_{(2)}^{cycl}(Q_1)n(Q_1) = \frac{1}{2}\gamma^2 N' - \gamma^2 n(Q_1), \quad (25)$$

where Eqs.(14) and (20) also are invoked. This result causes no surprise if we recall the additivity of any correction $E_{(k)}$ with respect to contributions of individual (cyclic and acyclic) pathways (i.e. the Rule 2 and Eq.(16)) and note that each ring $Q_1$ offers a separate local roundabout pathway of the second order. Moreover, the relation of Eq.(25) indicates that the above-observed destabilization of the $Q_1$-containing structures vs. the $Q_1$-free ones is entirely due to presence of just these simplest monocycles.

Consideration of underlying matrices $G_{(1)}$ (see the definition of $E_{(2)}$ of Eq.(9)) also supports the above conclusion: Zero elements stand in these matrices in the positions referring to cycles $Q_1$ (if any) in accordance with the vanishing matrix $G_{(1)}(Q_1)$ for an isolated two-membered ring $Q_1$ seen from Eq.(21). To illustrate this property, let us turn to Kekulè structures of the simplest hydrocarbon V (Fig. 4). Thus, matrices $G_{(1)}(V/II)$ and $G_{(1)}(V/III)$ representing the two $Q_1$-containing structures V/II and V/III are as follows

$$G_{(1)}(V/II) = -\frac{\gamma}{4}\begin{vmatrix} 0 & 1 & -1 & 0 \\ -1 & 0 & 1 & 0 \\ 1 & -1 & 0 & 0 \\ 0 & 0 & 0 & 0 \end{vmatrix}, \quad G_{(1)}(V/III) = -\frac{\gamma}{4}\begin{vmatrix} 0 & 1 & 0 & -1 \\ -1 & 0 & 1 & 0 \\ 0 & -1 & 0 & 0 \\ 1 & 0 & 0 & 0 \end{vmatrix} \quad (26)$$

and their elements take zero values in the positions 3,4 and 4,3. Meanwhile, the remaining matrix $G_{(1)}(V/I)$ of the only $Q_1$-free structure V/I contains more significant elements as exhibited below in Eq.(28).

Analogously, local roundabout pathways of the third order refer to individual monocycles $R_1$ and $S_1$ (if any) in the Kekulè structures concerned. Consequently, the third order corrections of Table 1 are additive quantities with respect to transferable contributions of these monocycles. Let the numbers of the latter to be correspondingly denoted by $n(R_1)$ and $n(S_1)$. The corrections $E_{(3)}$ then meet the following relation

$$E_{(3)} = E_{(3)}^{cycl}(R_1)n(R_1) + E_{(3)}^{cycl}(S_1)n(S_1) = \frac{3}{4}\gamma^3 n(R_1) - \frac{1}{4}\gamma^3 n(S_1) \quad (27)$$

[Absence of the acyclic component $E_{(3)}^{ac}$ should be recalled here]. As with Eq.(25), the above relation also may be easily rationalized by invoking the underlying matrices $G_{(1)}$ and $G_{(2)}$ (see Eq.(9)). For example, matrices $G_{(1)}(V/I)$ and $G_{(2)}(V/I)$ referring to the $R_1+S_1$-containing structure V/I of our model system V are as follows

$$G_{(1)}(V/I) = -\frac{\gamma}{4}\begin{vmatrix} 0 & 1 & -1 & 0 \\ -1 & 0 & 1 & 1 \\ 1 & -1 & 0 & -1 \\ 0 & -1 & 1 & 0 \end{vmatrix}, \quad G_{(2)}(V/I) = -\frac{\gamma^2}{8}\begin{vmatrix} 0 & 1 & -1 & 1 \\ -1 & 0 & 0 & 0 \\ 1 & 0 & 0 & 0 \\ -1 & 0 & 0 & 0 \end{vmatrix}. \quad (28)$$

It is seen that all off-diagonal elements of the matrix $G_{(1)}(V/I)$ take significant values except fot those referring to the most remote bonds $C_1=C_5$ and $C_4=C_8$. Meanwhile, the total number of non-zero elements is much lower in the second order matrix $G_{(2)}(V/I)$. In particular, the latter contains vanishing elements in the positions 2,3 and 3,2 so that no submatrix (block) like $G_{(2)}(R_1)$ arises there in spite of presence of the cycle $R_1$ in the structure concerned [For the matrix $G_{(2)}(R_1)$, see Eqs.(18) and (19)]. A more detailed analysis of expressions for elements $G_{(2)23}(V/I)$ shows that mediating increments of BOs of bonds $C_1=C_5$ and $C_4=C_8$ cancel out one another. This result causes little surprise if we recall (i) additivity of any element $G_{(2)il}$ with respect to mediators seen from Eq.(10) and (ii) opposite signs of mediating effects of BOs of endocyclic C=C bonds in the isolated monocycles $R_1$ and $S_1$ (Subsect. 3.2). The zero elements $G_{(2)23}(V/I)$, in turn, imply vanishing products $G_{(2)23} G_{(1)23}$ in the final formula for $E_{(3)}(V/I)$. The actual value of the latter then coincides with $2\gamma^3/4$ instead of $3\gamma^3/4$ representing an isolated ring $R_1$.

By contrast, the fourth order corrections of Table 1 ($E_{(4)}$) exhibit no additivity with respect to contributions of individual monocycles. The main reason for that is the more involved nature of the underlying pathways over BOs, namely these generally contain both cyclic and acyclic segments. Nevertheless, the destabilizing influence of the usual four-membered rings $Q_2$ (if any) is beyond any doubt. Indeed, the fourth order energies of the $Q_2$-containing structures (e.g. I/II, I/IV, II/I, etc.) are characterized by relatively low (negative) values vs. the remaining corrections $E_{(4)}$. In particular, the above-revealed lowest stability of the structure III/V of the bent isomer of naphthocyclobutene (III) (Subsect. 3.1) may be accounted for just by the destabilizing effect of the cycle $Q_2$. Again, the number of the latter ($n(Q_2)$) is unlikely to be the only parameter determining the fourth order energies, and the structures III/III and III/IV of the same isomer III may be invoked immediately to support this statement. Indeed, distinct values of corrections $E_{(4)}$(III/III) and $E_{(4)}$(III/IV) point to a certain role of more extended monocycles $R_2$ and/or $Q_3$ in their formation. An analogous conclusion follows also from comparison of fourth order energies $E_{(4)}$(IV/II) and $E_{(4)}$(IV/III) referring to the bent [3]phenylene (IV). These results are even more surprising if we recall that neither $R_2$ nor $Q_3$ may be entirely embraced by the fourth order energy $E_{(4)}$ as the Rule 3 of Section 2 indicates. A similar state of things has been found also in the case of Kekulè structures of benzenoids [35].

Finally, there are arguments for an important role of the four-membered cycle $S_2$ (if any) in the formation of corrections $E_{(4)}$. First, the actual value of the fourth order energy $E_{(4)}$(I/I) of the only $S_2$-containing Kekulè valence structure (I/I) of biphenylene (I) (equal to zero) proves to be relatively low as compared to the sum $2E_{(4)}(R_1) + E_{(4)}(Q_3)$ coinciding with $12\gamma^4/32$. Another argument follows from the constitution of the second order matrix $\boldsymbol{G}_{(2)}(I/I)$ exhibited below

$$\boldsymbol{G}_{(2)}(I/I) = -\frac{\gamma^2}{8} \begin{vmatrix} 0 & -1 & 1 & 0 & -1 & 0 \\ 1 & 0 & -1 & 1 & 0 & -1 \\ -1 & 1 & 0 & 0 & 1 & 0 \\ 0 & -1 & 0 & 0 & 1 & -1 \\ 1 & 0 & -1 & -1 & 0 & 1 \\ 0 & 1 & 0 & 1 & -1 & 0 \end{vmatrix}. \qquad (29)$$

It is seen that this matrix contains zero elements in the positions 1,4[4,1], 1,6[6,1], 3,4[4,3] and 3,6[6,3] referring to the substructure $S_2$ (Fig. 1) and this fact is in line with the vanishing matrix $\boldsymbol{G}_{(2)}(S_2)$ seen from Eq.(24). [Elements $G_{(2)13}(I/I)$ and $G_{(2)46}(I/I)$ (along with their counterparts in the positions 3,1 and 6,4, respectively) make an exception owing to non-zero mediating effects of BOs of the second and fifth double bonds].

On the whole, participation of the "non-standard" monocycles $S_1$ and $S_2$ in the formation of total $\pi$-electron energies of Kekulè valence structures of biphenylene-like hydrocarbons seems to be the most unexpected outcome of the above discussion. Thus, absence of these monocycles in the usual sets of conjugated circuits also may be among reasons of less satisfactory results of the theory of conjugated circuits when studying the REs of phenylenes [3, 45].

## 4. Conclusions

The above-suggested procedure to evaluate relative stabilities of individual Kekulè valence structures of a certain biphenylene-like hydrocarbon is based on a successive taking into account terms of power series for total $\pi$-electron energies ($E_{(k)}$) of increasing orders ($k$) (Subsect. 3.1). Again, the higher is the order parameter ($k$) of the given term $E_{(k)}$, the more extended monocycles participate in its formation (Subsect. 3.3). Consequently, a definite hierarchy of substructures (monocycles) reveals itself in respect of their importance when ordering Kekulè valence structures. In general, extinction of relative importance is observed when the size of the monocycle grows. Specific rules representing the same hierarchy of substructures are formulated below in terms of numbers of particular monocycles present in the Kekulè structures under comparison:

i) The number $n(Q_1)$ of the two-membered rings $Q_1$ is the principal parameter determining stabilities of Kekulè valence structures: The higher is this number, the lower is the relevant stability and vice versa;
ii) Given that two or several Kekulè valence structures are characterized by the same value of the above-specified parameter $n(Q_1)$, their relative stabilities depend on numbers $n(R_1)$ and $n(S_1)$ of the three-membered monocycles $R_1$ and $S_1$, respectively, viz. the higher is the former and the lower is the latter, the more stable is the structure concerned;
iii) For structures described by three uniform parameters $n(Q_1)$, $n(R_1)$ and $n(S_1)$, the relevant numbers of four- and five-membered monocycles start to play their role. In particular, the cycles $Q_2$ and $S_2$ contribute to destabilization of the given Kekulè valence structure.

Under an assumption of absence of even-membered monocycles in a certain Kekulè valence structure (as it is the case in those of benzenoid hydrocarbons), the above-enumerated rules (i)-(iii) automatically turn into the Fries rule, wherein the number $n(R_1)$ of benzenoid rings $R_1$ is the only decisive factor. Thus, the present rules may be regarded as an extension of the classical Fries rule to the case of biphenylene-like hydrocarbons.

Furthermore, the results obtained provide us with a deductive support for the principal assumptions underlying the simple intuition-based theory of conjugated circuits. In this respect, the above-concluded extinction of importance of a certain monocycle with its increasing size may be mentioned in the first place. Second, the usual monocycles (rings) $R_n$ and $Q_n$ (n=1,2…) consisting of C=C and C–C bonds alternately are now deductively shown to play the decisive role in ordering of Kekulè valence structures. Third, the above-specified rings $R_n$ and $Q_n$ are demonstrated to contribute to stabilization and to destabilization of the given Kekulè structure, respectively, as is usually supposed *a priori* on the basis of the classical 4n+2/4n rule [39, 40]. Finally, a sufficiently high extent of additivity of increments of individual monocycles follows from our perturbative expansion for total $\pi$-electron energies of Kekulè valence structures that also is among fundamentals of the theory of conjugated circuits. At the same time, the results of the present study point to the need for an extension of this theory when studying biphenylene-like hydrocarbons by including the monocycles $S_1$ and $S_2$ containing two and four exocyclic C=C bonds into the actual sets of the relevant conjugated circuits.

Significance of the results to theoretical and/or structural chemistry may be summarized as follows:
i) Distinctive features of biphenylene-like hydrocarbons vs. benzenoids are formulated explicitly by revealing additional structural factors (monocyclic substructures) that contribute to destabilization of some Kekulè structures of the former;
ii) Total $\pi$-electron energies of individual Kekulè valence structures are shown to offer an independent deductive criterion for evaluating their relative importances in the formation of the relevant actual electronic structures;
iii) Efficiency of the perturbational perspective on Kekulè valence structures is demonstrated, in particular for interpretation of their total $\pi$-electron energies in terms of contributions of specific substructures.

Table 1. Energy increments for Kekulè valence structures of hydrocarbons I-V along with the relevant total stabilization energies $(\Delta E)$ referring to particular values of the resonance parameter $\gamma$

| Structure | $E_{(2)}$ | $E_{(3)}$ | $E_{(4)}$ | $\Delta E(\gamma = 0.1)$ | $\Delta E(\gamma = 0.2)$ |
|---|---|---|---|---|---|
| I/I | $4\gamma^2$ | $6\gamma^3/4$ | 0 | 0.04150 | 0.1720 |
| I/II | $4\gamma^2$ | $5\gamma^3/4$ | $-14\gamma^4/32$ | 0.04121 | 0.1693 |
| I/III | $3\gamma^2$ | $6\gamma^3/4$ | $22\gamma^4/32$ | 0.03157 | 0.1331 |
| I/IV | $3\gamma^2$ | 0 | $-10\gamma^4/32$ | 0.02997 | 0.1195 |
| II/I | $4\gamma^2$ | $5\gamma^3/4$ | $-14\gamma^4/32$ | 0.04121 | 0.1693 |
| II/II | $4\gamma^2$ | $2\gamma^3/4$ | $8\gamma^4/32$ | 0.04053 | 0.1644 |
| II/III | $3\gamma^2$ | $3\gamma^3/4$ | $16\gamma^4/32$ | 0.03080 | 0.1268 |
| II/IV | $3\gamma^2$ | 0 | $-2\gamma^4/32$ | 0.02999 | 0.1199 |
| III/I | $4\gamma^2$ | $2\gamma^3/4$ | $-10\gamma^4/32$ | 0.04087 | 0.1635 |
| III/II | $3\gamma^2$ | $6\gamma^3/4$ | $14\gamma^4/32$ | 0.03154 | 0.1327 |
| III/III | $3\gamma^2$ | $3\gamma^3/4$ | $16\gamma^4/32$ | 0.03080 | 0.1268 |
| III/IV | $3\gamma^2$ | $3\gamma^3/4$ | $10\gamma^4/32$ | 0.03078 | 0.1265 |
| III/V | $3\gamma^2$ | $3\gamma^3/4$ | $-4\gamma^4/32$ | 0.03074 | 0.1258 |
| IV/I | $13\gamma^2/2$ | $9\gamma^3/4$ | $-\gamma^4/32$ | 0.06725 | 0.2780 |
| IV/II | $13\gamma^2/2$ | $7\gamma^3/4$ | $-29\gamma^4/32$ | 0.06666 | 0.2726 |
| IV/III | $13\gamma^2/2$ | $7\gamma^3/4$ | $-31\gamma^4/32$ | 0.06665 | 0.2725 |
| IV/IV | $9\gamma^2/2$ | $9\gamma^3/4$ | $41\gamma^4/32$ | 0.04738 | 0.2001 |
| V/I | $5\gamma^2/2$ | $2\gamma^3/4$ | $-17\gamma^4/32$ | 0.02585 | 0.1032 |
| V/II | $3\gamma^2/2$ | $3\gamma^3/4$ | $11\gamma^4/32$ | 0.01578 | 0.0666 |
| V/III | $3\gamma^2/2$ | 0 | $-7\gamma^4/32$ | 0.01498 | 0.0597 |

Table 2. Energy increments for isolated monocycles $R_n$, n=1,2,3,4, $Q_n$, n=1,2,3 and $S_n$, n=1,2
(N stands for the relevant number of C=C bonds)

| Monocycle | N | $E_{(0)}$ | $E_{(2)}$ | $E_{(3)}$ | $E_{(4)}$ |
|---|---|---|---|---|---|
| $R_1$ | 3 | 6 | $3\gamma^2/2$ | $3\gamma^3/4$ | $3\gamma^4/32$ |
| $R_2$ | 5 | 10 | $5\gamma^2/2$ | 0 | $5\gamma^4/32$ |
| $R_3$ | 7 | 14 | $7\gamma^2/2$ | 0 | $7\gamma^4/32$ |
| $R_4$ | 9 | 18 | $9\gamma^2/2$ | 0 | $9\gamma^4/32$ |
| $Q_1$ | 2 | 4 | 0 | 0 | 0 |
| $Q_2$ | 4 | 8 | $2\gamma^2$ | 0 | $-16\gamma^4/32$ |
| $Q_3$ | 6 | 12 | $3\gamma^2$ | 0 | $6\gamma^4/32$ |
| $S_1$ | 3 | 6 | $3\gamma^2/2$ | $-\gamma^3/4$ | $-5\gamma^4/32$ |
| $S_2$ | 4 | 8 | $2\gamma^2$ | 0 | $-16\gamma^4/32$ |